\documentclass[a4paper,11pt]{article}
\pdfoutput=1

\usepackage{jheppub}
\usepackage{amsmath}
\usepackage{siunitx}
\usepackage{graphicx}
\usepackage{multirow}
\usepackage{amssymb}
\usepackage{mathtools}
\usepackage{xcolor}
\usepackage{hyperref}
\usepackage{mathtools}
\usepackage{subcaption}
\usepackage[normalem]{ulem}

\catcode`\$=\active
\gdef$#1${\texorpdfstring{\(#1\)}{\detokenize{#1}}}

\newcommand{\ud}{\mathrm{d}}

\newcommand{\inmath}[1]{\relax\ifmmode#1\else$#1$\fi}

\begin{document}

\author[a,b]{Manoj K. Mandal,}
\emailAdd{mandal@pd.infn.it}

\author[a]{Xiaoran Zhao}
\emailAdd{xiaoran.zhao@uclouvain.be}
\affiliation[a]
{
    Centre for Cosmology, Particle Physics and Phenomenology (CP3),
    Universit\'{e} catholique de Louvain,
    1348 Louvain-la-Neuve,
    Belgium
}
\affiliation[b]
{
    Dipartimento di Fisica e Astronomia, Universit`a di Padova, Via Marzolo 8, 35131 Padova, Italy and INFN, Sezione di Padova, Via Marzolo 8, 35131 Padova, Italy
}

\title{Evaluating multi-loop Feynman integrals numerically through differential equations}

\abstract{The computation of Feynman integrals is often the bottleneck of multi-loop calculations. We propose and implement a new method  to efficiently evaluate such integrals in the physical region through the numerical integration of a suitable set of differential equations, where the initial conditions are provided in the unphysical region via the sector decomposition method. We present numerical results for a set of two-loop integrals, where the non-planar ones complete the master integrals for $gg\to\gamma\gamma$ and $q\bar{q}\to\gamma\gamma$ scattering mediated by the top quark.}

%\pacs{12.38.Bx,02.60.Lj}

\preprint{CP3-18-71, MCNET-18-32}

\maketitle

\section{Introduction}
With the advent of Run II of the Large Hadron Collider (LHC), a wealth of experimental 
measurements is expected to be performed at very high luminosities, probing the high energy scales extensively for the first time.  To exploit the full potential of these experimental 
measurements, theoretical predictions for the scattering processes are 
required with an unprecedented accuracy and precision. In several cases, the foreseen experimental precision will demand the inclusion of higher order terms in the perturbative expansions of the gauge coupling constants of the standard model, de facto requiring the evaluation of multi-loop amplitudes. Even though for processes
that can be mediated by heavy particles specific results have been obtained over the years~\cite{Graudenz:1992pv,Spira:1995rr,Baernreuther:2012ws,Czakon:2013goa,Borowka:2016ehy,Borowka:2016ypz,Jones:2018hbb,Baglio:2018lrj}, a general algorithm to efficiently, analytically and automatically compute the corresponding amplitudes is still lacking and poses an enormous challenge. As of now,  practical methods often rely on approximations~\cite{Bolzoni:2010xr,Cacciari:2015jma,Anastasiou:2015ema,Anastasiou:2016cez,Brucherseifer:2014ama,Berger:2016oht} and/or expansions\cite{Bonciani:2018omm,Davies:2018qvx}.

In general, a multi-loop amplitude can be expressed in terms of a finite set of integrals, usually known as master integrals. Although various methods for calculating the master integrals have been proposed (see Ref. \cite{Smirnov:2012gma} for a review), a fully general/universal one is not yet available (see Refs. \cite{Lee:2017qql,Liu:2017jxz,Borowka:2018dsa} for recent developments). However, the master integrals can be shown to satisfy differential equations~\cite{Kotikov:1990kg, Remiddi:1997ny, Gehrmann:1999as}, which after the reduction to a canonical form~\cite{Henn:2013pwa,Argeri:2014qva}, can be in some cases solved, iteratively. 
Although various results have been obtained in presence of the massive particles~\cite{Caron-Huot:2014lda,Bonciani:2016qxi,vonManteuffel:2017myy,Becchetti:2017abb,Mastrolia:2017pfy,Lee:2018rgs,Ablinger:2018yae,DiVita:2018nnh,Chen:2018fwb},
the final results are often represented as (iterated) integrals whose integrands consist of polylogarithms and other irrational functions, which still require numerical integration.

On the other hand, although solving differential equations numerically is a well-studied topic in applied mathematics, only a few phenomenological applications~\cite{Boughezal:2007ny,Czakon:2007qi,Czakon:2008zk} have been reported, till now. In such cases, the initial condition was obtained via expansions around a singular point, and finding out such expansions for other processes is highly non-trivial.

In the present work, we explore the possibility of evaluating Feynman integrals numerically
through differential equations, where the initial conditions are provided using
the sector decomposition method~\cite{Binoth:2000ps}.
The basic idea is simple: obtain the initial conditions in the unphysical region, which is a fast and accurate procedure, and then use the differential equations to analytically continue the results into the desired physical region. 

This work is organised as follows. In section \ref{sec:method} we describe the method in detail.
In section \ref{sec:examples} we illustrate the reach of our method by computing several two loop examples relevant for $gg\to\gamma\gamma$ and $q\bar{q}\to\gamma\gamma$ mediated by the top quark.  We draw our conclusions in section \ref{sec:conclusion}.

\section{Method}\label{sec:method}

We define a (scalar) Feynman integral In $d=4-2\epsilon$ dimensions by
\begin{equation}
    I=\left(\frac{e^{\epsilon\gamma_E}}{i\pi^{\frac{d}{2}}}\right)^{L}\int\prod_{i=1}^{L}\ud^d k_i\frac{1}{\prod_{j=1}^{N}D_j^{a_j}} ,\label{eq:definition}
\end{equation}
where $L$ is the number of loops, $k_i$ is the loop momentum, $N$ is the number of propagators and 
$D_j=q_j^2-m_j^2+i0^+$ is the denominator of the $j$-th propagator, where $q_j$ is the linear combination of the loop momenta and the external momenta, and $m_j$ is the corresponding mass. The $a_j$ denotes the respective power of the denominator.

The modern approach of multi-loop integrals consists in dividing the integrals  
into different topologies depending on their propagators. For each 
topology, a set of integration-by-parts (IBP) identities~\cite{Chetyrkin:1981qh}, relating different 
integrals, is generated exploiting the Poincar\'e invariance of the integrals.
With such system of linear identities at hand, any integral with the same 
topology can be written as a linear combination of a finite subset of integrals,
called the master integrals. Using the fact that derivatives of the master integrals with respect to the external kinematic variables and internal masses yield a linear combination of Feynman integrals in the same topologies,
IBP relations can be used to reduce them back to the linear combination of the master integrals,
leading to a system of first order partial differential equations.

Let us consider a vector of $M$ master integrals $I=(I_1,I_2,\cdots,I_M)^{\mathrm{T}}$, depending on $K$ independent kinematic variables $x=(x_1,x_2,\cdots,x_K)$ and $\epsilon$, one can express the set of equations as
\begin{equation}
    \frac{\partial I(x;\epsilon)}{\partial x_i}=J_i(x;\epsilon) I(x;\epsilon),~ ~ ~ i=1,\cdots,K \,,\label{eq:de}
\end{equation}
where $J_{i}$ is an $M\times M$ matrix, whose elements are rational functions of the kinematics $x$ and the dimension $d$. Each element of $J_{i}$ contains singularities originating from both the kinematics and the dimension $d$. The singularities from the kinematics are governed by the Landau equations~\cite{Landau:1959fi}, while the poles on $d$ must be rational numbers.

Although formally Eq.~\eqref{eq:de} is a set of partial differential equations, only one initial condition is needed to fix the solution and as a result such system can be integrated iteratively with respect to the kinematics, thereby making them similar to ordinary differential equations.
Therefore, the method for initial value problems~\cite{Stoer2002} can be applied straightforwardly to obtain the solution of the differential equation of the integrals. The main challenge is to obtain the suitable initial conditions  and design subsequent integration contours to fully fix the solution.

In the previous studies, an expansion around singular points~\cite{Czakon:2008zk,Lee:2017qql,Liu:2017jxz} was suggested.\footnote{In Ref. \cite{Caffo:2002ch},
    the results of the integral at singular points were adopted,
    which conflicts the Lipschitz condition and becomes ill-defined,
    thus requiring a modification,  which is equivalent to an expansion around singular points.}
However, such expansion is highly non-trivial,
and the short distance to singular points would lead to loss of accuracy and efficiency.\footnote{Here the loss of accuracy means the accuracy on the target points are much lower than the accuracy on the initial conditions.}

As the numerical algorithms are based on or related to the Taylor series expansion, 
the ideal initial conditions should be at the regular points,
far away from all the singularities. However, the computation of the integrals at those points by 
analytical or semi-analytical methods is as complicated as obtaining results at 
any regular points in the physical region.

In this work, we propose to obtain the initial conditions for the differential equations through 
the sector decomposition method~\cite{Binoth:2000ps}. All the ultraviolet 
and infrared divergences of the integrals are isolated in terms of a Laurent series in $\epsilon$, by dividing the 
integration domain and performing variable transformations according to well-designed strategies~\cite{Bogner:2007cr,Kaneko:2009qx}. The series can be expressed in the following form
\begin{equation}
    I=\sum_{i=0}^{+\infty}c_i\epsilon^{p+i},
    \label{eq:LaurentSeries}
\end{equation}
where $c_0$ represents the leading term, and the integer $p\in\mathbb{Z}$ is determined by the strength of the divergence of the integral.
The numerical values of the coefficients $c_i$ are obtained after performing a multi-dimensional integration.
In the unphysical region, where the $i0^{+}$ prescription is no longer needed,\footnote{Here we refer it as the unphysical region, but it also includes the physical region below all the thresholds of the internal particles so that the $i0^{+}$ prescription is not needed.} especially in the Euclidean region,
the integrands are sufficiently flat to achieve high precision through suitable multi-dimensional integration algorithm such as quasi-Monte Carlo algorithm~\cite{Li:2015foa}. At this point, one can exploit the analytic properties of the Feynman integrals:  considering the integral as a complex function,
the differential equations themselves can provide the analytical continuation
from the unphysical to the physical region.
As a consequence, the results of the integral in the physical region can be obtained as a Laurent series in $\epsilon$ as expressed in Eq.~\eqref{eq:LaurentSeries}.
On the other hand, with suitable contour deformations~\cite{Soper:1999xk,Anastasiou:2007qb},
the sector decomposition method can also provide the results for the physical kinematics.
Such a deformation, however,  requires a rather complicated variable transformation. 
In addition, the integrands still having large oscillations exhibit poor convergence in numerical integration.
Therefore, the direct computation via  sector decomposition in the physical region
tends to be computationally quite heavy.

An alternative path can be followed, by choosing the initial conditions in the unphysical region first and then by carefully choosing the contour of the integration.
To preserve the physical $i0^+$ prescription, the general idea is that along the contour except the target point, the integral do not require $i0^+$ prescription, and the target point is approached following the $i0^+$ prescription.
In general, constructing such contour is highly non-trivial,
and we give an example of a contour for those master integrals later in sec.\ref{sec:examples}. The contour is constructed carefully after the study of the branch cuts of the integrals and we leave the automation of the choice of the contours for the future.

As argued in the Ref. \cite{Czakon:2008zk}, 
explicit methods are sufficient to solve the system of differential equations.
They can be broadly organised into three classes: one-step (Runge-Kutta methods),
multi-step, and extrapolation methods. In practice, the final choice of a method in a specific problem depends on several criteria, including efficiency and availability.
In this work, we focus on one-step methods, mainly due to the following reasons:
\begin{enumerate}
    \item One-step methods only require one initial condition, in contrast with multi-step methods.\footnote{Some implementations of multi-step methods only apparently require one initial condition as one-step methods are used to provide other initial conditions.} This offers a great advantage since providing multiple initial conditions is a problem and may enhance uncertainty. In addition, it grants more freedom on the choice of the integration contour, as a piecewise contour can be adopted naively, e.g., the contour in section \ref{sec:examples}.
    \item The one-step methods are linear and with simple numerical coefficients.
        This yields negligible overhead time and very good numerical stability.
\end{enumerate}
We find that in order to achieve optimal efficiency, it is desirable to also introduce
an adaptive step-size control, as implemented in the Runge-Kutta-Fehlberg method.
In this method, for each step, two estimates of the results are obtained,
and the difference $\Delta I$ of them is calculated. Now, we have to define a relative error based on the $\Delta I$ and ${I}$, and then the adaptive step-size control is obtained through the comparison of this relative error to the desired local accuracy. 
Since $\Delta I$ and $I$ are Laurent series in $\epsilon$, the definition of the relative error is ambiguous. Now, for the purpose of defining a relative error, we observe the following facts. Firstly, as the integrals contribute in a non-trivial way to the evaluation of the amplitude, the uncertainty of each integral at the target point should be determined from the required precision on the value of the total amplitude itself. However, this determination can be very complicated in practical applications. Secondly, the uncertainties for the intermediate points should be based on the uncertainty of the final target point only, which is not known a priori, hence impossible to apply. Thirdly, we observe that while calculating the amplitude,
the uncertainties from different orders of $\epsilon$ usually mix together. Keeping these points in mind, we introduce the notion of the relative error of the integral, a quantity which is independent of any prefactor and based on the whole master integral rather than its individual terms in the $\epsilon$ expansion.
Considering a master integral $I$ with the difference $\Delta I$ described previously,
we define the relative error $\varepsilon_{\text{rel}}[\Delta I,I]$ based on the ratio $\frac{\Delta I}{I}$ as following:
\begin{align}
    \frac{\Delta I}{I}=&
    \frac
    {\sum_{i=0}^{n}\Delta c_i\epsilon^{i+p}+\mathcal{O}(\epsilon^{n+p+1})}
    {\sum_{i=0}^{n}c_i\epsilon^{i+p}+\mathcal{O}(\epsilon^{n+p+1})}
    =
    \sum_{i=0}^{n}b_i\epsilon^i+\mathcal{O}(\epsilon^{n+1}) \, , \\
    \varepsilon_{\text{rel}}\left[\Delta I, I\right]=&\max_{i}\left|b_i\right| .
\end{align}
And the maximum value of relative errors $\epsilon_{\textrm{rel}}[\Delta I, I]$ in the whole family is compared to the desired local accuracy, to control the step-size.

\section{Results}\label{sec:examples}
\begin{figure}[t]
    \centering
    \begin{subfigure}{0.23\textwidth}
        \includegraphics[width=\textwidth]{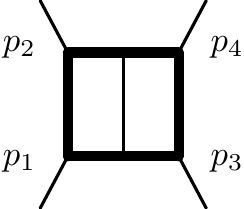}
        \caption{\label{fig:F1}$F_{1}$}
    \end{subfigure}
    \begin{subfigure}{0.23\textwidth}
        \includegraphics[width=\textwidth]{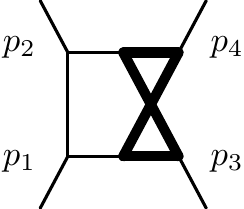}
        \caption{\label{fig:F2}$F_{2}$}
    \end{subfigure}
    \begin{subfigure}{0.23\textwidth}
        \includegraphics[width=\textwidth]{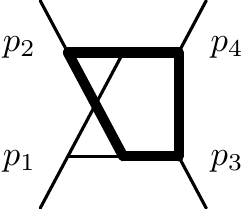}
        \caption{\label{fig:F3}$F_{3}$}
    \end{subfigure}
    \begin{subfigure}{0.23\textwidth}
        \includegraphics[width=\textwidth]{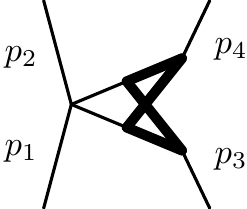}
		\caption{\label{fig:I2sub}$I_{2}^{\textrm{sub}}$}
    \end{subfigure}
	\caption{\label{fig:family}The three four-point two-loop integral families and $I_{2}^{sub}$ are shown here. $p_1,p_2$ are incoming and $p_3,p_4$ are outgoing. Thin lines represent massless particle, while thick lines are massive particles.}
\end{figure}

In the following, we demonstrate our method with three different 
planar and non-planar two-loop integral families, which appear in di-photon, di-jet production mediated by the heavy quarks. 
The diagrams are given in Fig. \ref{fig:family},
where $p_1,p_2$ are incoming and $p_3,p_4$ are outgoing.
The thin lines represent the massless particles, while the thick lines represent massive particles. 
All external lines are on-shell $p_1^2=p_2^2=p_3^2=p_4^2=0$,
and the kinematic variables are defined as $s=(p_1+p_2)^2,t=(p_1-p_3)^2,u=(p_1-p_4)^2$, which satisfy $s+t+u=0$. We normalise the invariants by the squared internal mass $m^2$,
effectively setting $m^2=1$, and the $m^2$ dependence can be recovered later, by power counting.

As explained before, for each master integral, we adopt \textsc{Nift}\cite{nift} to obtain the numerical results in the Euclidean region
by the sector decomposition method, where the final numerical integration is performed with the quasi-Monte Carlo algorithm.
We perform the IBP reduction with the C++ version of FIRE5~\cite{Smirnov:2014hma} together with 
LiteRed~\cite{Lee:2012cn,Lee:2013mka},
to obtain the corresponding differential equations in $s$ and $t$, treating them as independent variables. 
We perform the numerical integration of the differential equations with odeint~\cite{odeint},
and the Runge-Kutta-Fehlberg 7(8)-th order method~\cite{Fehlberg:68,Stoer2002} is chosen, based on our experimentation on one-loop integrals.

\begin{figure}
	\includegraphics[width=0.32\textwidth]{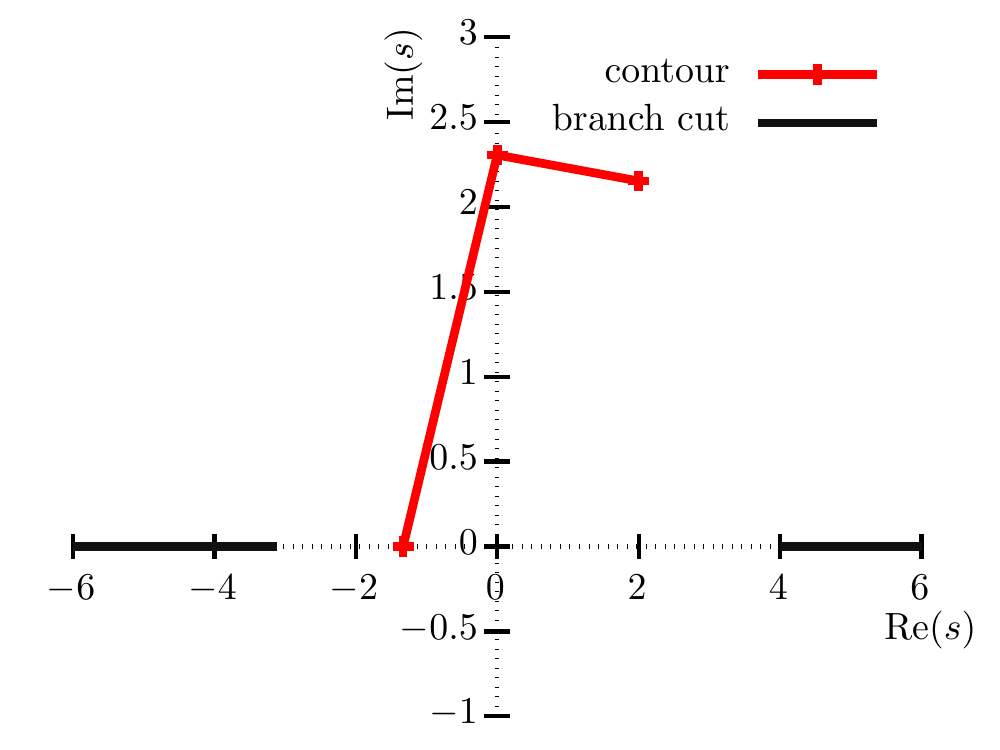}
	\includegraphics[width=0.32\textwidth]{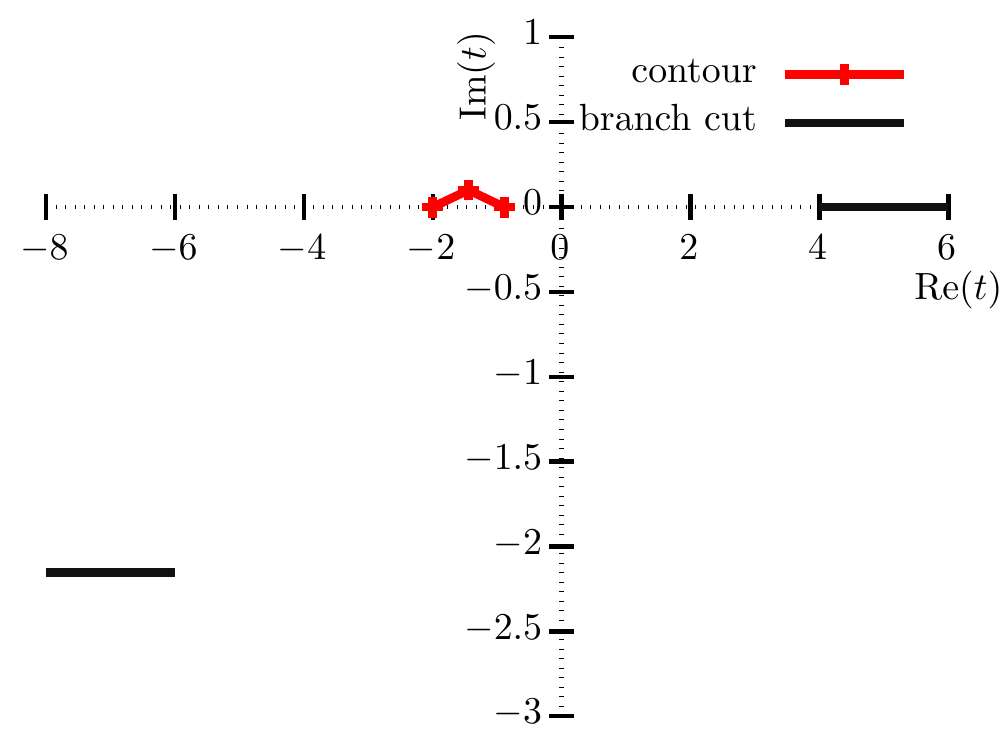}
	\includegraphics[width=0.32\textwidth]{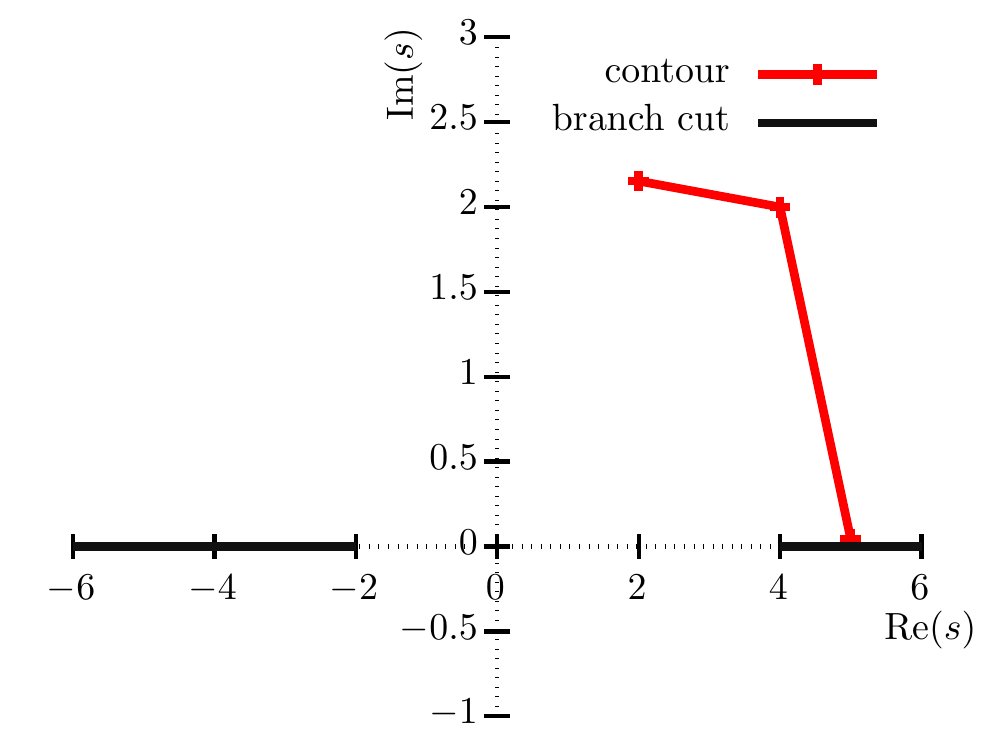}
	\caption{The integration contour (red) and relevant branch cuts (black) are shown for $F_3$, starting from IC1. Note that the branch cut corresponding to $u=4$ to $u=\infty$ is not present for $F_1$, and for $F_2$ one has an additional branch cut from $s=0$ to $s=4$.
	}\label{fig:branch-cut}
\end{figure}

We consider the target physical region defined by $s>4,t<0,u<0$,
and choose the initial conditions lying in the region with $s<0,t<0,u<4$.
We perform the evolution from the initial point $(s_a,t_a)$ to the target point $(s_b,t_b)$ along the following contour formed by six line segments:
\begin{align}
	\begin{split}
    &(s_a,t_a)\\
    \to&(i\sqrt{-4s_a},t_a)\\
    \to&(2+i\sqrt{-s_a}+i\sqrt{s_b-4},t_a)\\
    \to&(2+i\sqrt{-s_a}+i\sqrt{s_b-4},(t_a+t_b)/2+0.1i)\\
    \to&(2+i\sqrt{-s_a}+i\sqrt{s_b-4},t_b)\\
    \to&(4+i\sqrt{4(s_b-4)},t_b)\\
    \to&(s_b,t_b).
	\end{split}
\end{align}
In particular, we consider the target point with $(s,t)=(5,-2)$,
and we choose two different points in the Euclidean region as the initial points:
one is marked as IC1, with $(s,t)=(-1.33,-0.891)$;
another is marked as IC2, with $(s,t)=(-1.63,-0.632)$.
The difference between the results obtained from those two different initial conditions provides an estimate of the uncertainties.
We list all branch points on the physical Riemann sheet in table \ref{tab:bp}, and we verified that the above contour never crosses branch cut,
as can be seen in fig. \ref{fig:branch-cut}.
Alternatively, instead of determining the branch points and the branch cuts,
along the contour the sector decomposition method can be adopted to calculate the numerical values of the Feynman integrals directly,
since we require that along the contour the $i0^+$ prescription is not needed.
Such numerical values provide another cross check on the results obtained from the numerical integration of differential equations.

\begin{table}
    \centering
    \begin{tabular}{c|c|c|c}
        \hline
        & F1 & F2 & F3 \\
        \hline
        $s=0$ & N & Y & N \\
        $s=4$ & Y & Y & Y \\
        $s=-16$ & - & N & - \\
        $t=0$ & N & N & N\\
        $t=4$ & Y & Y & Y \\
        $u=0$ & N & N & N\\
        $u=4$ & - & Y & Y \\
        $t=u$ & - & N & N \\
		$st+4u=0$ & Y/N & Y/N & Y/N\\
        $tu+4s=0$ & - & Y/N & Y/N \\
        $su+4t=0$ & - & Y/N & Y/N \\
        $4t^2-s(t-1)^2=0$ & N & N & N \\
		$4u^2-s(u-1)^2=0$ & - & - & N
    \end{tabular}
    \caption{The full list of singularities other than infinity is shown,
    as well as whether it is a branching point(marked as "Y") or not(marked as "N"). If such point is not a singular point of corresponding family, "-" is shown.
Note that we adopt $u=-s-t$ to show the crossing symmetry. For $st+4u=0$ it becomes a branching point only when $s>0,t>0$, thus we mark it as "Y/N", similarly for the other two $tu+4s=0$ and $su+4t=0$.\label{tab:bp}}
\end{table}

All the timings reported here are based on a laptop with Intel Core i5-6200U 
CPU and the time cost consists of the evaluation of all the master integrals in 
the whole family. We require the relative error on the initial conditions less than $10^{-7}$, and 
the relative error tolerance in each step of the differential equations is set to $10^{-10}$.

We begin with the family $F_1$, where the analytical results in $d=4$ dimension have been reported in Ref.~\cite{Caron-Huot:2014lda}.
We choose the denominators as\footnote{Technically, to perform the IBP reduction, two extra denominators should be chosen. However, we choose all master integrals to be scalar master integrals without any numerator, hence the results are independent of the exact form of the auxiliary denominators in the IBP reduction. We neglect the two extra denominators here for simplicity. The above comment also applies to the other two families.}:
\begin{align}
	\begin{split}
    D_1=k_1^2-m^2,
    D_2=(k_1-p_1)^2-m^2,
    D_3=(k_1-p_1-p_2)^2,\\
    D_4=k_2^2-m^2,
    D_5=(k_2-p_3)^2-m^2,
    D_6=(k_2-p_1-p_2)^2-m^2,\\
    D_7=(k_1-k_2)^2.
	\end{split}
\end{align}
We denote the integrals in this family as $I(F_1,a_1a_2a_3a_4a_5a_6a_7)$,
where $a_i$ is the corresponding propagator power, as described in Eq. ~\eqref{eq:definition}.
Working in $d=4-2\epsilon$ dimension, after the IBP reduction, we obtain 29 master integrals.
In Table \ref{tab:comp-ic}, we show the initial conditions of one of the top level master integrals $I_1=I(F_1,1111111)$.
As mentioned before, the relative uncertainty on the initial conditions are required to be less than $10^{-7}$,
and our results are consistent with analytical ones\cite{Caron-Huot:2014lda} within such uncertainty.
Using those two initial conditions,
we evaluate these integrals for the benchmark value($s=5, t=-2$) in the physical region,
and the results of $I_1$ are shown in Table \ref{tab:comp-ana}.
We also report the numerical value obtained from the analytical expression in Ref.~\cite{Caron-Huot:2014lda}.
We find that the uncertainty of our numerical results compared to the analytical one is less than $10^{-6}$. Moreover, the difference between the results obtained using the initial conditions from IC1 and IC2 is also of the same order, providing a good estimate on the uncertainty. We note that to reach such high precision takes only 0.1s.

\begin{table}[t]
    \centering
    \begin{tabular}{cccccc}
        & & & $c_0$ & time(s)\\
        \hline
		\multirow{4}{*}{$I_1$} & \multirow{2}{*}{IC1} & \textsc{Nift} & $-0.059087788(6)$ & 1.93  \\
                               & & Ref. \cite{Caron-Huot:2014lda} & $-0.059087788\phantom{(0)}$ & -- \\\cline{2-5}
							   & \multirow{2}{*}{IC2} & \textsc{Nift} & $-0.056016652(5)$ & 1.74\\
                               & & Ref. \cite{Caron-Huot:2014lda} & $-0.056016650\phantom{(0)}$ & --\\
        \hline
		\multirow{4}{*}{$I_2^{\text{sub}}$} & \multirow{2}{*}{IC1} & \textsc{Nift} & $0.28729542(1)$ & 3.55 \\
                                            & & Ref. \cite{vonManteuffel:2017hms} & $0.28729543\phantom{(0)}$ & -- \\\cline{2-5}
											& \multirow{2}{*}{IC2} & \textsc{Nift} & $0.26181028(1)$ & 3.57 \\
                                            & & Ref. \cite{vonManteuffel:2017hms} & $0.26181029\phantom{(0)}$ & -- \\
    \end{tabular}
    \caption{\label{tab:comp-ic}The comparison of our numerical initial conditions obtained from \textsc{Nift}\cite{nift} with the analytical ones for the Feynman integral $I_1$ and $I_2^{\text{sub}}$. The two initial points are: IC1($s=-1.33,t=-0.891$) and IC2($s=-1.63,t=-0.632$). $c_0$ is the leading term of the $\epsilon$ expansion of these finite integrals.}
\end{table}

\begin{table}[t]
    \centering
    \begin{tabular}{cccc}
        \multicolumn{2}{c}{$(s=5, t=-2)$}  & $c_0$  &  time(s) \\
        \hline
        \multirow{3}{*}{$I_1$} & IC1 &  $\phantom{+}0.573661717-i0.45602298$ & 0.11 \\
                               & IC2 & $\phantom{+}0.573662051-i0.45602316$ & 0.10 \\
                               & Ref. \cite{Caron-Huot:2014lda} & $\phantom{+}0.573661756-i0.45602309$ & -- \\
        \hline
        \multirow{3}{*}{$I_2^{\text{sub}}$} &IC1 & $-0.077764616+i0.34306744$ & 0.26 \\
                                            & IC2 & $-0.077764595+i0.34306737$ & 0.23 \\
                                            & Ref. \cite{vonManteuffel:2017hms} & $-0.077764620+i0.34306741$ & -- \\
    \end{tabular}
    \caption{\label{tab:comp-ana}The comparison of our numerical results with the analytical ones for the Feynman integral $I_1$ and $I_2^{\text{sub}}$ at the point $(s=5, t=-2)$. The IC1 and IC2 denotes the two different choices of the initial conditions. $c_0$ is the leading term of the $\epsilon$ expansion of these finite integrals.}
\end{table}
The next example is the family $F_2$, shown in 
Fig. \ref{fig:F2}, with the following denominators:
\begin{align}
	\begin{split}
    D_1=k_1^2,
    D_2=(k_1-p_1)^2,
    D_3=(k_1-p_1-p_2)^2,\\
    D_4=k_2^2-m^2,
    D_5=(k_2-p_1-p_2+p_3)^2-m^2,\\
    D_6=(k_1-k_2)^2-m^2,
    D_7=(k_1-k_2-p_3)^2-m^2 .
	\end{split}
\end{align}
There are 36 master integrals in this family, and some of them involve infrared divergences. 
The most complicated integrals in this family, i.e. the seven-propagator master integrals, are still unknown in literature \footnote{Partial results has been reported in Ref.~\cite{Xu:2018eos} recently.}.
Instead, for comparison, we show numerical results for one non-planar integral in the lower sector, defined by $I_2^{\text{sub}}=I(F_2,1011111)$\footnote{An alternative numerical evaluation for this topology has been reported in Ref.~\cite{Bonciani:2018uvv}} (shown in fig. \ref{fig:I2sub}),
which has been studied in Ref.~\cite{vonManteuffel:2017hms}
and in fact  is independent of $t$.
In Table \ref{tab:comp-ic}, we show our numerical initial conditions obtained from \textsc{Nift} as well as the analytical one on $I_2^{\text{sub}}$.
The uncertainties on the initial conditions are less than $10^{-7}$ and the computing time is well under control.
\begin{table}[t]
    \centering
    \footnotesize
    \setlength\tabcolsep{2pt}
    \begin{tabular}{cccccc}
        \multicolumn{2}{c}{$(s=5, t=-2)$} & $c_0$ & $c_1$ & $c_2$ & time(s) \\
        \hline
        \multirow{3}{*}{$I_2$} & IC1 & $0.02188084-i0.00000002$ & $-0.0870259+i0.05170117$ & $-0.246416-i0.17602070$ & 0.26\\
                               & IC2 & $0.02188080+i0.00000001$ & $-0.0870262+i0.05170118$ & $-0.246417-i0.17602072$ & 0.23\\
                               & pySecDec & $0.02187(3)\ +i0.00003(3)\ $ & $-0.0869(3)\ +i0.0518(4)\ \ \ $ & $-0.248(2)\ -i0.175(2)\ \ \ \ $ & $\mathcal{O}(10^4)$\\
        \hline
        \multirow{3}{*}{$I_3$} & IC1 & $-0.0599222+i0.4204527$ & $-1.2093294+i1.1271787$ & $-3.737851+i0.435880$ & 0.74\\
                               & IC2 & $-0.0599219+i0.4204528$ & $-1.2093298+i1.1271798$ & $-3.737851+i0.435879$ & 0.78\\
                               & pySecDec & $-0.05998(7)+i0.42048(8)$ & $-1.2100(7)\ +i1.1262(7)\ $ & $-3.737(3)\ +i0.430(3)\ $ & $\mathcal{O}(10^4)$ \\
    \end{tabular}
    \caption{\label{tab:comp-new} Comparison between numerical results obtained with our algorithm from two differential choices of initial conditions
        for the Feynman integral $I_2$ and $I_3$ at the point $(s=5, t=-2)$.
        $c_0, c_1$ and $c_2$ denotes the first three coefficients in the Laurent series of $\epsilon$.
    The results obtained from pySecDec~\cite{Borowka:2017idc} is also shown for consistency check and the corresponding setup is not optimal. }
\end{table}
In Table \ref{tab:comp-ana}, we show our numerical results as well as the analytical one on $I_2^{\text{sub}}$ in the physical region with $s=5$.
Similarly to $I_1$, the uncertainty from our approach is less than $10^{-6}$.
The time cost is several times larger than F1, but still less than 1 second. At the same time, we also obtain the results for the seven-propagator integral $I_2=I(F_2,1111111)$. As no analytical results are known for this, we use pySecDec~\cite{Borowka:2017idc} to obtain the results for cross check.
In table~\ref{tab:comp-new} we show the results for all the coefficients starting from $\epsilon^{-2}$ to $\epsilon^{0}$ in $\epsilon$ expansion. 
By estimating the uncertainties of our method through the difference between the two results,
the relative error is at $\mathcal{O}(10^{-6})$. This is much more accurate
than directly evaluating it via the sector decomposition method in the physical region.

Finally, we consider family $F_3$, shown in Fig. \ref{fig:F3}. This family \footnote{Results in the Euclidean region has been reported recently in Ref.~\cite{Xu:2018eos}.} contains 51 master integrals, and in particular five of them belong to the seven-propagator sector, indicating more complicated differential equations than the family $F_1$ and $F_2$.
The propagators are given by:
\begin{align}
	\begin{split}
    D_1=k_1^2,D_2=(k_1-p_1)^2,\\
    D_3=k_2^2-m^2,
    D_4=(k_2-p_4)^2-m^2,
    D_5=(k_2-p_3-p_4)^2-m^2,\\
    D_6=(k_1-k_2)^2-m^2,
    D_7=(k_1-k_2+p_2)^2-m^2.
	\end{split}
\end{align}
We use the same points IC1 and IC2 to obtain the initial conditions.
We show the numerical results for $I_3=I(F_3,1111111)$ in the Table \ref{tab:comp-new}
and further checked with pySecDec.
The computing cost of our method is still less than one second, and the precision of our results is still at $\mathcal{O}(10^{-6})$.

The computing cost on multi-dimensional integration for obtaining the initial conditions varies from 17 
seconds to 2 minutes depending on the complexity, which is much less than 
the time spent for IBP reduction, hence negligible in practical application.
The number of steps for the numerical integration of the differential equations ranges from 61 to 133, thereby indicating that the discretisation error associated with the differential equations is at most around $10^{-8}$.
As explained before, the dominant uncertainties come from the uncertainties on the initial conditions,
and we verified it by adjusting the relative error tolerance on the initial conditions and/or the differential equations.
Further information, including numerical results for all master integrals in the three integral families are available as ancillary files with the arXiv submission.

Remarkably, one does not need to start from the Euclidean region each time.
Once the results at one physical point are obtained according to previous procedure,
they can be adopted as the new initial condition, for other physical points.
As the branch points and branch cuts in the physical region are well-understood,
comparing to the general cases,
much simpler contours can be adopted.

\section{Conclusion and discussion}\label{sec:conclusion}
In this paper, we have presented a method to compute the Feynman integrals numerically.
The main idea is to integrate the differential equations numerically,
with the initial condition in the Euclidean region provided through the sector decomposition method.
We have compared numerical results achieved by our method with the available analytical ones,
for several two-loop examples, 
and shown $\mathcal{O}(10^{-6})$ accuracy can be reached within one second.
Using the above method, we have provided numerical results of several two-loop integrals, whose analytical expressions are currently unknown.
Those two-loop integrals complete the two-loop master integrals for $gg\to\gamma\gamma$ and $q\bar{q}\to\gamma\gamma$ scattering mediated by the top quark,
and thus our results can be applied directly to investigate the role of top quark in di-photon production at the LHC.

The differential equations of the integral encode the full $\epsilon$ dependence, while the sector decomposition can provide any higher order terms in $\epsilon$. Therefore, the results of the integral at any order of $\epsilon$ expansion can be achieved within our method, which is usually desirable and required in practical applications.

Although in this paper we restricted it to the case with real masses only,
our method can be applied to the case with complex masses.
Clearly, the sector decomposition method works with complex masses.
On the other hand, the integration contour still doesn't cross any branch cut if the width is small.
Such complex mass scheme, is crucial and essential to describe 
the threshold behaviour for processes involving unstable particles.
In that case, we hope our method will provide an important role to obtain the precise prediction of relevant processes.

Our method builds up on the idea that the differential equations of the master integrals can be integrated  numerically, providing an initial condition in the unphysical region and a suitable integration contour.

However, it is not always possible to obtain the differential equations of the master integrals as the IBP reduction usually fails in case of the integrals having a large number of scales. 
In this context, for example, the recent proposal~\cite{Mastrolia:2018uzb, Frellesvig:2019kgj} of the use of intersection theory could overcome this problem.
%The recent proposal of the use of intersection theory~\cite{Mastrolia:2018uzb} could be an alternative of the IBP reduction.
The initial conditions are obtained by using the sector decomposition method,
which is quite efficient in the Euclidean region.
While it can be a problem for massless cases,
for processes with massive loop propagators, usually such region can be found.

Finally, we note that as both the IBP reduction and sector decomposition can be done systematically
and automatically, our method could play a potential role towards an automated approach and framework to multi-loop computations. This, of course, only if an algorithm for the automatic determination of the integration contours could be identified. Work in this direction is in progress.

\begin{acknowledgments}
    We thank Fabio Maltoni and Pierpaolo Mastrolia for useful discussions and comments on manuscript. The authors would also like to thank Johannes Henn, Roman N. Lee and Andreas von Manteuffel for discussions and providing additional numerical results on cross checks.
    XZ has received funding from the European Union's Horizon 2020 research and innovation programme as part of the Marie Sk\l{}odowska-Curie Innovative Training Network MCnetITN3 (grant agreement no. 722104). MKM has been supported by the Research Fellowship grant awarded by the Belgian Federal Science Policy Office (BELSPO) and by the UniPD STARS Grant 2017 ``Diagrammalgebra".
\end{acknowledgments}

\appendix

\bibliographystyle{JHEP}
\bibliography{numi.bib}

\end{document}